\documentclass{webofc}
\usepackage[varg]{txfonts}   % Web of Conferences font
\usepackage{bm}
\usepackage{amsmath}
\usepackage{graphicx}
\usepackage[bookmarksnumbered=true,
bookmarksopen=true,
colorlinks=true,
pdfborder={0 0 1},
citecolor=red,
linkcolor=blue,
anchorcolor=green,
urlcolor=blue,
breaklinks=true]{hyperref}

\begin{document}
\title{Synchronization of Singularly Perturbed Systems with Time Scales}
%
% subtitle is optionnal
%
%%%\subtitle{Do you have a subtitle?\\ If so, write it here}

\author{\firstname{Neng} \lastname{Wan}\inst{1}\fnsep\thanks{\email{wanxx179@d.umn.edu}} \and
        \firstname{Desineni S.} \lastname{Naidu}\inst{2}\fnsep\thanks{\email{dsnaidu@d.umn.edu}} 
        % etc.
}

\institute{Department of Mathematics and Statistics, University of Minnesota Duluth, Duluth, MN 55812, USA
\and
          Department of Electrical Engineering, University of Minnesota Duluth, Duluth, MN 55812, USA
}

\abstract{%
Synchronization problems of continuous and discrete singularly perturbed systems are studied in this paper with singular perturbations and time scales (SPaTS) technique. The dynamics of leader and followers are decomposed into pure-slow and pure-fast subsystems. Locally optimal decentralized tracking sub-controllers are synthesized respectively to asymptotically synchronize each subsystem of follower. A composite control protocol is proposed to synchronize the original dynamics of each follower with the leader's dynamics. Both analogous and digital flight control systems for aircraft formation flying are utilized to verify the effectiveness of the control schemes. 
}
\maketitle
\section{Introduction}
\label{sec1}
With more and more distributed systems emerged, such as unmanned aerial or ground vehicle (UAV or UGV) teams, sensor and robot networks, the area of decentralized control has grown to be significantly active during the last decade. Consensus and synchronization are two of the most popular problems in decentralized control, which also referred as distributed control~\cite{Negenborn_CSM_2014, Wang_TAC_2013}, coordinated control~\cite{Cao_TII_2013}, or cooperative control~\cite{Ren_2008} in some references. Among most of the existing literatures, consensus was referred as the task that the agents with distinct initial states ultimately reach a certain common agreement through a local decentralized control protocol without a leader~\cite{Saber_PIEEE_2007}, and the synchronization was referred as the behaviour of enforcing all agents’ states to follow the state of a leader~\cite{Lu_TNNLS_2012}. More detailed introduction and recent progress on decentralized control are available in the survey papers~\cite{Cao_TII_2013, Saber_PIEEE_2007} and books~\cite{Ren_2008, Lewis_2013}. People have studied the decentralized control algorithm for many different systems, such as the first-, second- and higher-order linear systems~\cite{Su_TAC_2016}, partial differential equation systems~\cite{Pilloni_TAC_2016}, complex and nonlinear systems~\cite{Tong_TSMC_2016}. However, the decentralized control for a kind of interesting model, singularly perturbed systems with slow and fast time scales, has rarely been considered so far. 

Singularly perturbed  systems, differential equations systems with small parameters multiplying certain derivatives, commonly exist in our physical systems due to the presence of some parasitic parameters, such as resistance, inductances, capacitances in electronic systems, small time constants, moment of inertial, and Reynolds number in the mechanical systems~\cite{Naidu_JGCD_2001, Gajic_2001}. Singular perturbations and time scales (SPaTS), which decomposes the singularly perturbed systems into pure-slow and pure-fast subsystems and designs the sub-controllers respectively, is an attractive technique when designing the controller for singularly perturbed systems. SPaTS reduces the order of plant model by decomposing the original system into two subsystems and allows different control methods to be employed to stabilize each subsystems. Some recommended survey papers on SPaTS are~\cite{Naidu_JGCD_2001, Naidu_2002, Zhang_IJISS_2014}. Although there exist plenty of research papers on SPaTS underlying both the continuous-time~\cite{Zhang_WSEAS-TSC_2014} and discrete-time control~\cite{Litkouhi_TAC_1985, Zhang_IJSS_2016} systems, the existing works seldom investigate the decentralized control problem with SPaTS, which is one of the thrusts of this research paper. 

In this research paper both the continuous- and discrete-type of SPaTS techniques are revisited and are adopted to synthesize the synchronization controllers for decentralized control systems with time scales. Chang transformation~\cite{Chang_SIAM-JMA_1972} is utilized to decompose the original systems into slow and fast subsystems. For each subsystem, an optimal synchronization controller is synthesized subject to a local performance criterion. An aircraft control system is used to verify the effectiveness of this control scheme and simulate the formation flying of the aircraft networks in practice. The remainder of this paper is organized as follows. \hyperref[sec2]{Section 2} formulates the control problem and introduces some preliminary knowledge on graph theory. \hyperref[sec3]{Section 3} and \hyperref[sec4]{Section 4} respectively synthesize the synchronization controllers for continuous-time and discrete-time control systems. \hyperref[sec5]{Section 5} verifies the control schemes with aircraft control systems for formation flying. \hyperref[sec6]{Section 6} draws the conclusions.

The notations used throughout this paper are illustrated as follows. For a matrix $\bm{M}$, $\bm {M}^T$ is its transpose; $\bm{M}^*$ is its conjugate transpose; and $\sigma_{\max} (\bm{M})=\sqrt{\lambda_{\max}(\bm{M}^*\bm{M})}$ represents its largest singular value. For a complex number $c$, $\textrm{Re}(c)$ represents its real part.

\section{Problem Formulation}\label{sec2}
In this section, both the continuous- and discrete-type singularly perturbed models are given and interpreted. Basic concepts in graph theory are briefly introduced and used to describe the communication networks between agents. Synchronization control problems are formulated in the end of this section.

\subsection{Singularly Perturbed Models}
In general, after some nonsingular transformations, most of the continuous-time linear systems with slow- and fast-time scales can be formulated as follows
\begin{equation}\label{eq1}
	\begin{cases}
		\dot{\bm x}_1(t)  = \bm{A}_1 \bm{x}_1(t) + \bm{A}_2\bm{x}_2(t)+\bm{B}_1\bm{u}(t)\\
		\dot{\bm x}_2(t)  = \epsilon^{-1} \bm{A}_3\bm{x}_1(t) + \epsilon^{-1}\bm{A}_4\bm{x}_2(t) + \epsilon^{-1}\bm{B}_2\bm{u}(t)\\
	\end{cases}
\end{equation}
where $\bm x_i(t) \in \mathbb{R}^{n_i}$ for $i=1,2$; $\bm{u}(t) \in \mathbb{R}^m$; $\epsilon$ is a small positive parameter; the fast subsystem matrix $\bm A_4 \in \mathbb{R}^{n_2 \times n_2}$ is nonsingular; and the dimensions of other matrices in~\eqref{eq1} can be obtained accordingly. Multiplying each side of the second equation in~\eqref{eq1} by $\epsilon$, the standard form of a continuous-time singularly perturbed linear time-invariant system is given as follows
\begin{equation}\label{eq2}
	\begin{cases}
		\dot{\bm x}_1(t)  = \bm{A}_1 \bm{x}_1(t) + \bm{A}_2\bm{x}_2(t)+\bm{B}_1\bm{u}(t)\\
		\epsilon\dot{\bm x}_2(t)  = \bm{A}_3\bm{x}_1(t) + \bm{A}_4\bm{x}_2(t) + \bm{B}_2\bm{u}(t)\\
	\end{cases}
\end{equation}
which is usually referred as the model expressed in a slow time scale~\cite{Gajic_2001}. 

Sampling~\eqref{eq1} with a sampling period $T$ or~\eqref{eq2} with a sampling period $\epsilon T$, the discrete-time singularly perturbed systems in a fast time scale can be described by the following difference equation
\begin{equation}\label{eq3}
	\begin{cases}
		\bm{x}_1(k+1) = \left(\bm{I} + \epsilon\bm{A}_1\right)\bm{x}_1(k) + \epsilon\bm{A}_2\bm{x}_2(k) + \epsilon\bm{B}_1\bm{u}(k)\\
		\bm{x}_2(k+1) = \bm{A}_3\bm{x}_1(k) + \bm{A}_4\bm{x}_2(k) + \bm{B}_2\bm{u}(k)\\
	\end{cases}
\end{equation}
where $\bm x_i(k) \in \mathbb{R}^{n_i}$ for $i=1,2$; $\bm u (k) \in \mathbb R^m$; $\epsilon$ is a small positive parameter; the fast subsystem matrix $\bm I- \bm A_4$ is nonsingular; and the dimensions of other matrices in~\eqref{eq3} can be told accordingly. It is worth noting that although identical symbols are used in~\eqref{eq2} and~\eqref{eq3} for convenience, for a same system in continuous and discrete cases, matrices $\bm A_i$ and $\bm B_i$ are usually assigned with different values in~\eqref{eq2} and~\eqref{eq3}, for example, the aircraft control system discussed in~\hyperref[sec5]{Section 5}.

\subsection{Graph Theory and Communication Network}
Some basic notations in graph theory are introduced in the following lines to describe the communication topology among agents. Let $\bm{\mathcal{N}}=\left\{1,2,\cdots,N\right\}$ be the index set of followers. We use a graph $\bm{\mathcal{G}}=\left\{\bm{\mathcal{V}}, \bm{\mathcal{E}} \right\}$ to represents the communication network, where the vertex set $\bm{\mathcal{V}}=\left\{v_1,v_2,\cdots,v_N\right\}$ is used to represent the agents or followers, and the edge set $\bm{\mathcal{E}} \subset \bm{\mathcal{V}}\times \bm{\mathcal{V}}$ is used to describe the communication channels between agents. An edge $(v_j,v_i) \in \bm{\mathcal{E}}$ from vertex $j$ to vertex $i$ means that agent $i$ can sense the states from agent $j$. Define $\bm{\mathcal{A}} = [\alpha_{ij}] \in \mathbb{R}^{N \times N}$ be the adjacency matrix associated with graph $\bm{\mathcal{G}}$, and let $\alpha_{ii}=0$ and $\alpha_{ij}>0$ if $(v_j, v_i) \in \bm{\mathcal{E}}$; otherwise, $\alpha_{ij}=0$. The Laplacian matrix associated with graph $\bm{\mathcal{G}}$ is $\bm{\mathcal{L}}=[l_{ij}]$, where $l_{ij}=-\alpha_{ij}$ if $i\neq j$, and $l_{ij}=\sum_{k=1}^{N}\alpha_{ik}$  if $i=j$. 

\subsection{Control Problem}
In this paper, both the continuous and discrete synchronization or decentralized tracking controllers are synthesized with SPaTS technique. As it was supposed in~\cite{Zhang_TAC_2011}, we assumes that the leader or the reference model in continuous problem has the following dynamics
\begin{equation}\label{eq4}
	\begin{cases}
		\dot{\bm{x}}_1^{(0)}(t) = \bm{A}_1 \bm{x}_1^{(0)}(t) + \bm{A}_2 \bm{x}_2^{(0)}(t)\\
		\dot{\bm{x}}_2^{(0)}(t) = \epsilon^{-1}\bm{A}_3\bm{x}_1^{(0)}(t) + \epsilon^{-1}\bm{A}_4\bm{x}_2^{(0)}(t)\\
	\end{cases}
\end{equation}\label{eq5}
which is model~\eqref{eq1} without control input. According to~\cite{Lewis_2013, Zhang_TAC_2011}, the autonomous model~\eqref{eq4} is able to generate a large class of useful command trajectories. Correspondingly, the leader or the reference model in discrete case has the dynamics as follows
\begin{equation}
	\begin{cases}
	\bm{x}_1^{(0)}(k+1) = \left(\bm{I} + \epsilon\bm{A}_1\right)\bm{x}_1^{(0)}(k) + \epsilon\bm{A}_2 \bm{x}_2^{(0)}(k)\\
	\bm{x}_2^{(0)}(k+1) = \bm{A}_3\bm{x}_1^{(0)}(k) + \bm{A}_4\bm{x}_2^{(0)}(k) 
	\end{cases}
\end{equation}
The objective of synchronization or decentralized tracking control problem within the preceding settings can be formulated as to design a local decentralized control protocol $\bm u^{(i)}(t)$ or $\bm u^{(i)}(k)$ such that $\lim_{t\rightarrow\infty}[\bm{x}^{(i)}(t)- \bm{x}^{(0)}(t)]=\mathbf{0}$ or $\lim_{k\rightarrow\infty}[\bm x^{(i)}(k)-\bm x^{(0)}(k)]=\mathbf{0}$ for $\forall i \in \bm{\mathcal{N}}$, where the superscript $(i)$ identifies the state or control of different agents. Meanwhile, we assume that there always exists a directed path from leader to every follower node.

\begin{figure}[htpb]
	\centering
	{
		\includegraphics[width=0.65\textwidth]{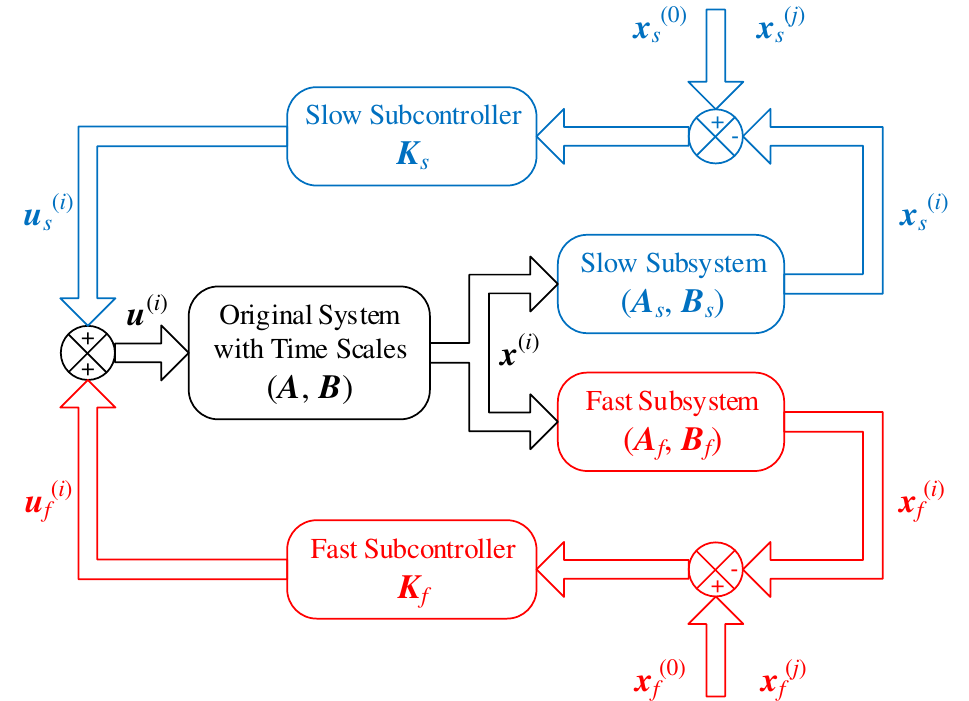}
		\caption{\label{fig1} Architecture of a local synchronization controller.}
	 }
\end{figure}

By applying SPaTS technique to design the synchronization controller, original systems~\eqref{eq2} and~\eqref{eq3} with time scales are decomposed into pure-slow and pure-fast subsystems. Local decentralized sub-controllers $\bm u_s^{(i)}$ and $\bm u_f^{(i)}$ are separately designed for the slow and fast subsystems. The sub-controllers make up the overall synchronization controller $\bm u^{(i)}$. \hyperref[fig1]{Figure 1} illustrates the control architecture of a local synchronization controller designed by SPaTS technique.

\section{Continuous Synchronization Controller}\label{sec3}
In this section, synchronization controller for the continuous systems~\eqref{eq2} is investigated with SPaTS technique. Original system~\eqref{eq2} is decomposed into two subsystems with different time scales via Chang transformation, and then locally optimal sub-controllers are designed for each subsystems. 

\subsection{Model Decomposition}
Consider the continuous singularly perturbed system~\eqref{eq2}, and the superscript $(i)$ is omitted in this subsection for conciseness. In order to decompose~\eqref{eq2} into two subsystems with different time scales, define the fast state vector as
\begin{equation}\label{eq6}
	\bm x_f(t) = \bm{x}_2(t) + \bm{Mx}_1(t)
\end{equation}
where $\bm x_f(t) \in \mathbb{R}^{n_2}$ and $\bm M \in \mathbb{R}^{n_2 \times n_1}$. Multiplying each side of~\eqref{eq6} by $\epsilon$ and taking the first derivative with respect to time, we have
\begin{equation}\label{eq7}
	\epsilon \dot{\bm{x}}_f = \left(\bm{A}_3 + \epsilon\bm{MA}_1 - \bm{A}_4\bm{M} - \epsilon\bm{MA}_2\bm{M} \right)\bm{x}_1(t) + \left(\bm{A}_4 + \epsilon\bm{MA}_2\right)\bm{x}_f(t) + \left(\bm{B}_2+\epsilon\bm{MB}_1\right)\bm{u}(t)
\end{equation}
By letting the coefficient before $\bm{x}_1(t)$ be zero, we obtain the decoupled pure-fast dynamics
\begin{equation}\label{eq8}
	\bm{x}_f(t)= \bm{A}_f\bm{x}_f(t)+\bm{B}_f\bm{u}(t)
\end{equation}
where $\bm{A}_f= \epsilon^{-1}\left(\bm{A}_4 +\epsilon\bm{MA}_2\right)$ and $\bm{B}_f=\epsilon^{-1}\left(\bm{B}_2+\epsilon\bm{MB}_1\right)$, and an algebraic equation
\begin{equation}\label{eq9}
	\bm{A}_3 + \epsilon \bm{MA}_1 - \bm{A}_4\bm{M} - \epsilon\bm{MA}_2\bm{M} = \mathbf{0} .
\end{equation}
In order to find the decoupled pure-slow dynamics, define the slow state vector as follows
\begin{equation}\label{eq10}
	\bm{x}_s(t) = \bm{x}_1(t) - \bm{Nx}_f(t) ,
\end{equation}
where $\bm{x}_s(t)\in\mathbb{R}^{n_1}$ and $\bm{N} \in \mathbb{R}^{n_1\times n_2}$. Take the first derivative of~\eqref{eq10} with respect to time, we have
\begin{equation}\label{eq11}
	\begin{split}
		\dot{\bm{x}}_s(t) =& \left(\bm{A}_1\bm{N} + \bm{A}_2 - \bm{A}_2\bm{MN} + \bm{NMA}_2 - \epsilon^{-1}\bm{NA}_4\right)\bm{x}_f(t) \\
		&\hspace{20pt} + \left(\bm{A}_1 - \bm{A}_2\bm{M}\right)\bm{x}_s(t) + \left(\bm{B}_1-\epsilon^{-1}\bm{NB}_f\right)\bm{u}(t)
	\end{split}
\end{equation}
By letting the coefficient before $\bm{x}_f(t)$ be zero, the decoupled pure-slow dynamics is obtained
\begin{equation}\label{eq12}
	\dot{\bm{x}}_s(t) = \bm{A}_s\bm{x}_s(t) + \bm{B}_s\bm{u}(t) 
\end{equation}
where $\bm{A}_s=\bm{A}_1-\bm{A}_2\bm{M}$, $\bm{B}_s=\bm{B}_2+\epsilon\bm{MB}_1$, and the algebraic equation
\begin{equation}\label{eq13}
	\bm{A}_1\bm{N} + \bm{A}_2 - \bm{A}_2\bm{MN} + \bm{NM}\bm{A}_2 - \epsilon^{-1}\bm{NA}_4 = \mathbf{0}
\end{equation}
We can solve the matrices $\bm{M}$ and $\bm{N}$ from equations~\eqref{eq9} and~\eqref{eq13}.

With the decoupled subsystems~\eqref{eq8} and~\eqref{eq12}, local synchronization control protocol $\bm u_s (t)$ and $\bm u_f(t)$ are separately designed for each subsystem. According to \hyperref[fig1]{Figure 1}, the overall controller of the original system should have the form of $\bm u(t) = \bm u_s(t)+ \bm u_f(t)$.

\subsection{Synchronization Protocol}
Based on the transformations~\eqref{eq6} and~\eqref{eq10}, the reference signals for fast and slow subsystems are correspondingly $\bm x_f^{(0)}(t)=\bm x_2^{(0)}(t)+\bm{Mx}_1^{(0)}(t)$ and $\bm x_s^{(0)}(t)=\bm x_1^{(0)} (t)-\bm{Nx}_f^{(0)}(t)$. Since the slow and fast subsystems have same formulation when omitting the subscripts $f$ and $s$, we synthesize the synchronization protocols simultaneously on dynamics
\begin{equation}\label{eq14}
	\dot{\bm{x}}(t) = \bm{Ax}(t) + \bm{Bu}(t) .
\end{equation}

Define the neighborhood tracking error of node $i$ has the following form
\begin{equation}\label{eq15}
	\bm{e}^{(i)}(t) = \sum_{j=1}^{N}\alpha_{ij}\left(\bm{x}^{(j)} - \bm{x}^{(i)}\right) + \beta_i\left(\bm{x}^{(0)} - \bm{x}^{(i)}\right)
\end{equation}
where $\alpha_{ij}$ is the element in adjacency matrix $\bm A$ and $\beta_i$ is the pinning gain. We define the pinning gain $\beta_i>0$ if node $i$ can measure the states of leader, otherwise $\beta_i=0$; and we define the matrix $\bm{\mathcal{B}}=diag(\beta_1,\beta_2,\cdots,\beta_n)$. Design the decentralized synchronization controller as follows
\begin{equation}\label{eq16}
	\bm{u}^{(i)}(t) = c\bm{K}\bm{e}^{(i)}(t).
\end{equation}
Synchronization can be achieved or the tracking errors $\bm x^{(i)}(t)-\bm x^{(0)}(t)$ will asymptotically converge to zeros when the coupling gain
\begin{equation}\label{eq17}
	c\geq \frac{1}{2 \min_{i\in\bm{\mathcal{N}}}{\textrm{Re}(\lambda_i)}}
\end{equation}
with $\lambda_i$ the eigenvalues of matrix $\bm{\mathcal{L}}+\bm{\mathcal{B}}$~\cite{Lewis_2013}. In order to minimize the local performance index
\begin{equation}\label{eq18}
	J^{(i)} = \frac{1}{2}\int_{0}^{\infty}\left\{[\bm{x}^{(i)}(t)]^{T}\bm{Q}\bm{x}^{(i)}(t)+[\bm{u}^{(i)}(t)]^{T}\bm{R}\bm{u}^{(i)}(t)\right\} dt
\end{equation}
with decentralized controller~\eqref{eq16}, choose the control gain matrix
\begin{equation}\label{eq19}
	\bm{K}= \bm{R}^{-1}\bm{B}^T\bm{P} ,
\end{equation}
where matrices $\bm{Q} \in\mathbb{R}^{n_i\times n_i}$ and $\bm{R} \in \mathbb{R}^{m\times m}$ are positive definite, and the positive definite matrix $\bm{P} \in \mathbb{R}^{n_i \times n_i}$ is the unique solution of the following algebraic Riccati equation
\begin{equation}\label{eq20}
	\bm{A}^{T}\bm{P} + \bm{PA} + \bm{Q} - \bm{PBR}^{-1}\bm{B}^T\bm{P} = \mathbf{0}
\end{equation}
Although the performance index~\eqref{eq18} is defined with the transformed states~\eqref{eq6} and~\eqref{eq10}, the overall performance index can be minimized by either separating the original performance index before designing the sub-controllers or properly assign the gain matrices $\bm{Q}$ and $\bm{R}$ when designing the sub-controllers, which has been comprehensively discussed in~\cite{Naidu_NASA_1988, Naidu_1988}. After we designed the sub-controllers $\bm{u}_f^{(i)}(t)$ and $\bm{u}_s^{(i)}(t)$ with~\eqref{eq8} and~\eqref{eq12} respectively, the composite controller for the original system~\eqref{eq1} or~\eqref{eq2} of agent $i$ can be formulated as
\begin{equation}\label{eq21}
	\bm{u}^{(i)}(t)=\bm{u}^{(i)}_s+\bm{u}^{(i)}_f = c\left(\bm{K}_s\bm{e}_s^{(i)} + \bm{K}_f\bm{e}_f^{(i)}\right)
\end{equation}
Consequently, the closed-loop dynamics of agent $i$ becomes
\begin{equation}\label{eq22}
	\dot{\bm{x}}^{(i)}(t) = \bm{Ax}^{(i)}(t) + c\bm{B}\left[\bm{K}_s\bm{e}_s^{(i)}(t) + \bm{K}_f\bm{e}_f^{(i)}(t)\right] ,
\end{equation}
where
\begin{equation*}
	\bm{x}^{(i)}(t) = 
	\left[
	\begin{matrix}
	\bm{x}_1^{(i)}(t)\\
	\bm{x}_2^{(i)}(t)\\
	\end{matrix}
	\right], 
	\quad
	\bm{A} = 
	\left[
	\begin{matrix}
	\bm{A}_1 & \bm{A}_2\\
	\epsilon^{-1}\bm{A}_3 & \epsilon^{-1}\bm{A}_4\\
	\end{matrix}
	\right], 	
	\quad
	\bm{B}=
	\left[
	\begin{matrix}
	\bm{B}_1\\
	\epsilon^{-1}\bm{B}_2\\
	\end{matrix}
	\right]
\end{equation*}

\section{Discrete Synchronization Controller}\label{sec4}
Discrete systems are more popular nowadays, since digital controllers are widely deployed around us. In this section, synchronization controller for the discrete system~\eqref{eq3} is investigated with SPaTS. Discrete dynamics~\eqref{eq3} is decomposed into two subsystems with different time scales, and locally optimal synchronization control protocols are separately synthesized for each subsystems.

\subsection{Model Decomposition}
Consider the discrete singularly perturbed system~\eqref{eq3}. Omitting the superscript $(i)$, define the fast state vector as
\begin{equation}\label{eq23}
	\bm{x}_f(k) = \bm{x}_2(k)+ \bm{Mx}_1(k) . 
\end{equation}
Substituting~\eqref{eq3} into~\eqref{eq23}, we have
\begin{equation}\label{eq24}
\begin{split}
	\bm{x}_f(k+1) = &\left(\epsilon\bm{MA}_2 + \bm{A}_4\right)\bm{x}_f(k) + \left(\bm{MB}_1 + \bm{B}_2\right)\bm{u}(k) \\
	&\hspace{20pt}+ \left(\bm{M}+\epsilon\bm{MA}_1 -\epsilon\bm{MA}_2\bm{M} + \bm{A}_3 - \bm{A}_4\bm{M}\right)\bm{x}_1(k)
\end{split}
\end{equation}
By letting the coefficient before $\bm x_1(k)$ be zero, we have the decoupled pure-fast dynamics
\begin{equation}\label{eq25}
	\bm{x}_f(k+1) = \bm{A}_f\bm{x}_f(k) + \bm{B}_f\bm{u}(k)
\end{equation}
where $\bm{A}_f=\epsilon\bm{MA}_2 + \bm{A}_4$ and $\bm{B}_f=\bm{MB}_1+\bm{B}_2$, and the algebraic equation
\begin{equation}\label{eq26}
	\bm{M} + \epsilon\bm{MA} - \epsilon\bm{MA}_2\bm{M} + \bm{A}_3 - \bm{A}_4\bm{M} = \mathbf{0} . 
\end{equation}
Define the slow state vector as
\begin{equation}\label{eq27}
	\bm{x}_s(k) = \bm{x}_1(k) - \epsilon\bm{Nx}_f(k)
\end{equation}
Then by substituting~\eqref{eq3} and~\eqref{eq25} into~\eqref{eq27}, we have
\begin{equation}\label{eq28}
	\begin{split}
		\bm{x}_s(k+1) &= \left(\epsilon\bm{N} + \epsilon^2 \bm{A}_1\bm{N} + \epsilon\bm{A}_2 - \epsilon^2 \bm{A}_2\bm{MN} - \epsilon\bm{NA}_f\right) \bm{x}_f(k)\\
		&\hspace{30pt} + \left(\bm{I} + \epsilon\bm{A}_1 - \epsilon\bm{A}_2\bm{M}\right)\bm{x}_s(k) + \left(\bm{B}_1 - \epsilon\bm{NB}_f\right)\bm{u}(k)
	\end{split}
\end{equation}
By letting the coefficient before $\bm x_f(k)$ be zero, the pure-slow dynamics becomes
\begin{equation}\label{eq29}
	\bm{x}_s(k+1) = \bm{A}_s\bm{x}_s(k) + \bm{B}_s\bm{u}(k)
\end{equation}
where $\bm{A}_s=\bm{I}+\epsilon\bm{A}_1-\epsilon\bm{A}_2\bm{M}$ and $\bm B_s=\bm B_1-\epsilon\bm{NB}_f$. An algebraic equation is obtained
\begin{equation}\label{eq30}
	\epsilon\bm{N} + \epsilon^2\bm{A}_1\bm{N} + \epsilon\bm{A}_2 - \epsilon^{2}\bm{A}_2\bm{MN} - \epsilon\bm{NA}_f = \mathbf{0}
\end{equation}
Matrices $\bm{M}$ and $\bm{N}$ can be solved from algebraic equations~\eqref{eq26} and~\eqref{eq30}, which will be discussed later in~\hyperref[sec5]{Section 5}. 

\subsection{Synchronization Protocol}
According to the transformations~\eqref{eq23} and~\eqref{eq27}, the reference signals of two subsystems or the fast and slow states of leader are $\bm x_f^{(0)}(k)= \bm x_2^{(0)}(k)+ \bm{Mx}_1^{(0)}(k)$ and $\bm{x}_s^{(0)}(k)=\bm{x}_1^{(0)}(k) - \epsilon\bm{Nx}_f^{(0)}(k)$. Omitting the subscript $f$ and $s$ in~\eqref{eq25} and~\eqref{eq29}, we investigate the discrete synchronization protocols on the following dynamics 
\begin{equation}\label{eq31}
	\bm{x}(k+1) = \bm{Ax}(k) + \bm{Bu}(k) .
\end{equation}

The neighborhood tracking error of node $i$ has the same form as~\eqref{eq15}. The discrete decentralized synchronization controller of follower $i$ is designed as follows
\begin{equation}\label{eq32}
	\bm{u}_i(k) = c\left(1+d_i+\beta_i\right)^{-1}\bm{Ke}_i(k) ,
\end{equation}
where $d_i=l_{ii}$ is the diagonal element in Laplacian matrix $\bm{\mathcal{L}}$, $\beta_i$ is the pinning gain. Define $\bm{\mathcal{D}}=diag(d_1,d_2,\cdots,d_n)$. Local performance index  
\begin{equation}\label{eq33}
	J^{(i)} = \frac{1}{2} \sum_{k=0}^{\infty}\left[\bm{x}^{(i)}(k)\right]^{T}\bm{Qx}^{(i)}(k) + \left[\bm{u}^{(i)}(k)\right]^T\bm{R}\rho\bm{u}^{(i)}(k)
\end{equation}
in the limit as $\rho \rightarrow 0$ can be minimized if the feedback gain matrix
\begin{equation}\label{eq34}
	\bm{K} = \left(\bm{B}^T\bm{PB}\right)^{-1}\bm{B}^T\bm{PA}
\end{equation}
where matrices $\bm{Q} \in \mathbb{R}^{n_i \times n_i}$ and $\bm{R} \in \mathbb{R}^{m \times m}$ are positive definite, and the positive definite matrix $\bm P \in \mathbb{R}^{n_i \times n_i}$ is a solution of the discrete-time Riccati-like equation
\begin{equation}\label{eq35}
	\bm{A}^T\bm{PA} - \bm{P} + \bm{Q} - \bm{A}^T\bm{PB}\left(\bm{B}^T\bm{PB} \right)^{-1}\bm{B}^{T}\bm{PA} = \mathbf{0}
\end{equation}
Define the weighted graph matrix 
\begin{equation}\label{eq36}
	\bm{\Gamma} = \left(\bm{I} + \bm{\mathcal{D}} + \bm{\mathcal{B}}\right)^{-1}\left(\bm{\mathcal{L}} + \bm{\mathcal{B}} \right)
\end{equation}
and the raidus
\begin{equation}\label{eq37}
	r = \left[\sigma_{\max}\left(\bm{Q}^{-T/2}\bm{A}^T\bm{PB}\left(\bm{B}^T\bm{PB}\right)^{-1}\bm{B}^T\bm{PAQ}^{-1/2}\right)\right]^{-1/2}
\end{equation}
Synchronization can be achieved or the tracking errors $\bm x^{(i)}(k) - \bm x^{(0)}(k)$ will asymptotically converge to zeros when the coupling gain $c$ satisfies
\begin{equation}\label{eq38}
	cr_0 < r
\end{equation}
where all the eigenvalues of weighted graph matrix $\bm{\Gamma}$ are located inside the covering circle centered at $(1/c,0)$ with radius $r_0$. Since we can tell that the coupling gain $c$ is not determined by the communication topology only, after we solve the sub-controllers $\bm u_f^{(i)}(k)$ and $\bm u_s^{(i)}(k)$, the composite controller for the original discrete system~\eqref{eq3} of agent $i$ is designed as
\begin{equation}\label{eq39}
	\bm{u}^{(i)}(k) = \left(1+d_i + \beta_i\right)^{-1}\left(c_s\bm{K}_s\bm{e}_s^{(i)}+c_f\bm{K}_f\bm{e}_f^{(i)}\right)
\end{equation}
Therefore, the closed-loop dynamics of agent $i$ becomes
\begin{equation}\label{eq40}
	\bm{x}^{(i)}(k+1) = \bm{Ax}^{(i)}(k) + \bm{B}\left(1+d_i + \beta_i\right)^{-1}\left(c_s\bm{K_s}\bm{e}_s^{(i)} + c_f\bm{K_f}\bm{e}_f^{(i)}\right) .
\end{equation}

\section{Numerical Simulation}\label{sec5}
In this section, we verify the synchronization controllers proposed in the preceding sections with both continuous and discrete flight control systems to simulate the scenario of aircraft formation flying in practice.

The communication network among aircrafts is shown in~\hyperref[fig 2]{Figure 2}. The leader agent 0 can either be a physical aircraft or a virtual command generator. Both aircraft 1 and 2 can sense the command signal from the leader, and the states of aircraft 1 can be measured by the aircrafts 2 and 3.
\begin{figure}[htpb]
	\centering{
		\includegraphics[width=0.4\textwidth]{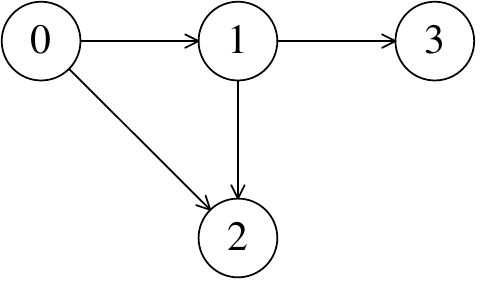}
		\caption{\label{fig2} Communication network between aircrafts.}
	}
\end{figure}
The adjacency matrix $\bm{\mathcal{A}}$ and the Lagrangian matrix $\bm{\mathcal{L}}$ associated with the graph in~\hyperref[fig 2]{Figure 2} are
\begin{equation*}
	\begin{split}
	\bm{\mathcal{A}} = 
	\left[\begin{matrix}
	0&0&0\\
	1&0&0\\
	1&0&0\\
	\end{matrix}\right]
	\textrm{\quad and \quad}
		\bm{\mathcal{L}} = 
		\left[\begin{matrix}
		0&0&0\\
		-1&1&0\\
		-1&0&1\\
		\end{matrix}\right]	
	\end{split}
\end{equation*}
$\bm{\mathcal{B}} = diag\left(1,1,0\right)$, and $\bm{\mathcal{D}} = diag\left(0, 1, 1\right)$.

\subsection{Continuous Flight Control System }
\begin{figure}[htpb]
	\centering
	\includegraphics[width=0.58\textwidth]{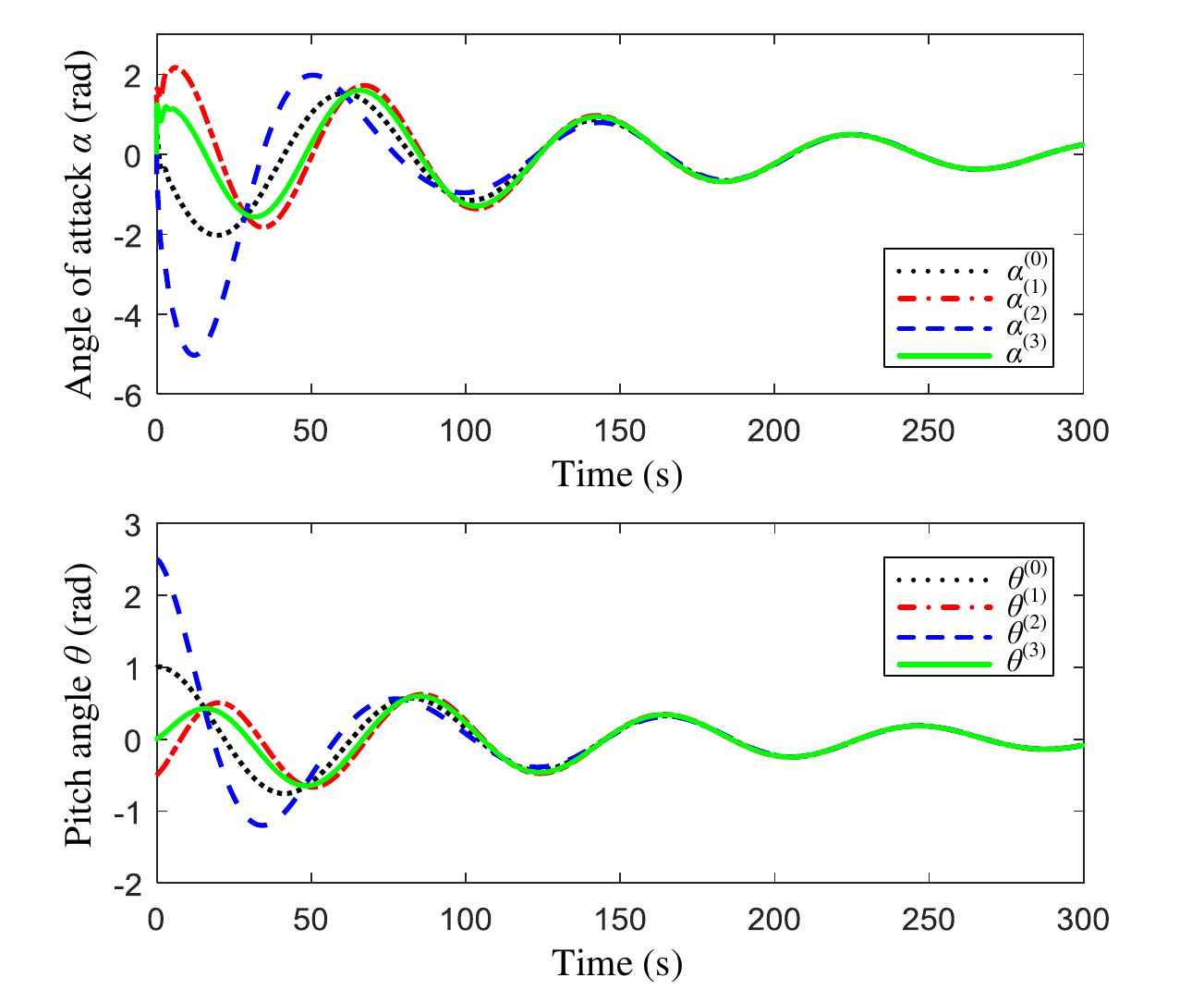}
	\caption{\label{fig3} Longitudinal attitude of aircrafts.}
\end{figure}

\begin{figure}[htpb]
	\centering
	\includegraphics[width=0.58\textwidth]{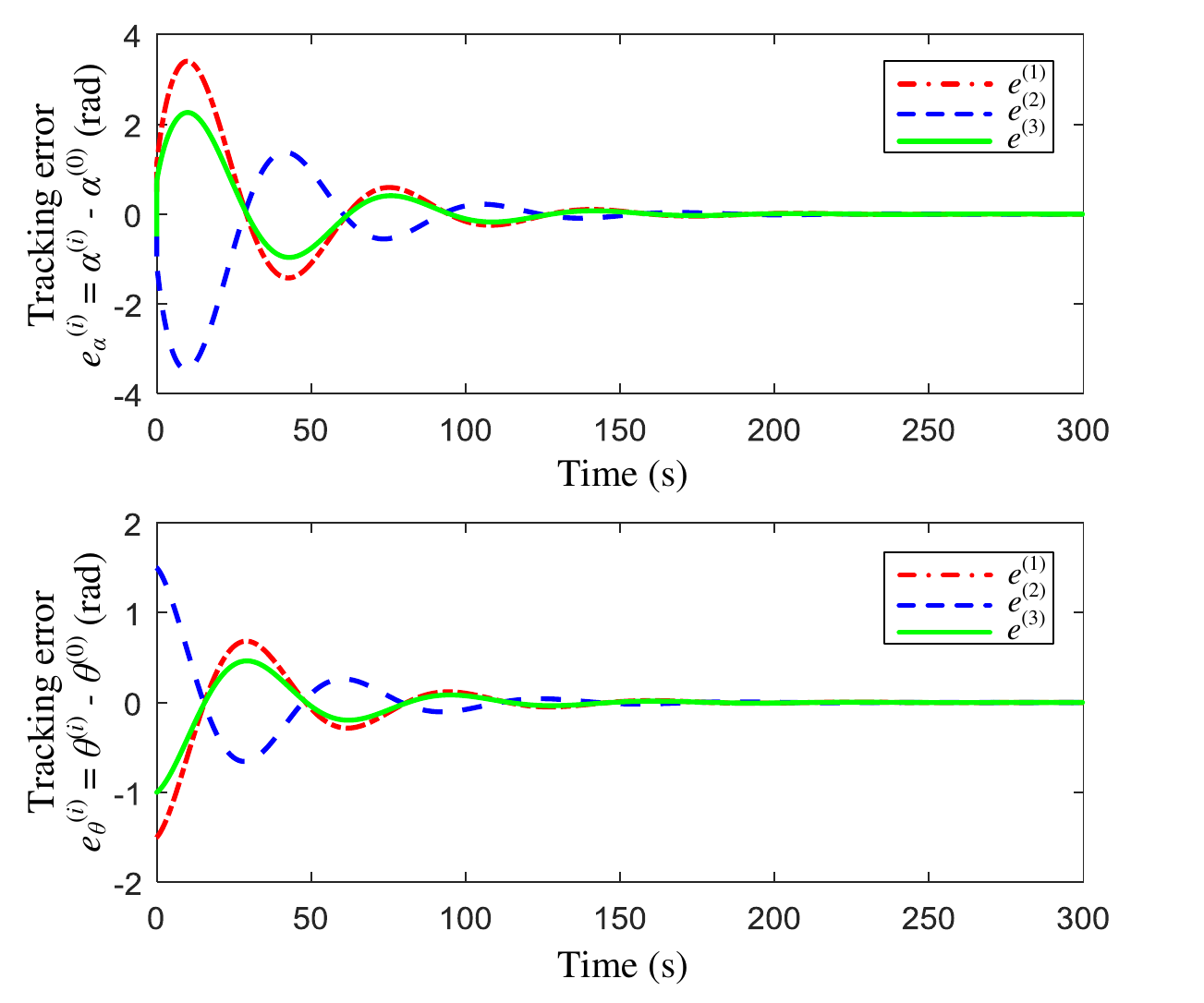}
	\caption{\label{fig4} Tracking error of longitudinal attitude.}
\end{figure}

The continuous dynamics for the longitudinal motion control of an aircraft is given as follows~\cite{Elliott_TAC_1977}:
\begin{equation}\label{eq41}
	\dot{\bm{x}}(t) = \bm{A}_c\bm{x}(t) + \bm{B}_c\bm{u}(t)
\end{equation}
where
\begin{equation*}
	\begin{split}
		\bm{A}_c = \left[\begin{matrix}
		-0.015 & -0.0805 & -0.0011666 & 0\\
		0 & 0 & 0 & 0.03333\\
		-2.28 & 0 & -0.84 & 1\\
		0.6 & 0 & -4.8 & -0.49\\
		\end{matrix}\right]
		\textrm{, }
		\bm{B}_c^{T} = \left[\begin{matrix}
		-9.16 \times 10^{-4} & 0 &-0.11 & -8.7\\
		7.416 \times 10^{-4} & 0 & 0 & 0\\
		\end{matrix}\right] .
	\end{split}
\end{equation*}
$\bm x(t)=[u, \theta, q, \alpha]^T$, $u(t)=[\delta_e,\delta_T]^T$, and $n_1=n_2=2$. Matrices $\bm M$ and $\bm N$ in~\eqref{eq6} and~\eqref{eq10} can be solved from equations~\eqref{eq9} and~\eqref{eq13} by using the following Newton algorithm:
\begin{equation}\label{eq42}
	\bm{E}^{[i]}_1\bm{M}^{[i+1]} + \bm{M}^{[i+1]}\bm{E}_2^{[i]} = \bm{F}^{[i]}
\end{equation}
where superscript $[i]$ represents the times of iteration $\bm E_1^{[i]} = \bm A_4+ \epsilon \bm{M}^{[i]}\bm{A}_2$, $\bm{E}_2^{[i]}=-\epsilon(\bm{A}_1-\bm{A}_2\bm{M}^{[i]})$, $\bm{F}^{[i]} =\bm{A}_3+\epsilon\bm{M}^{[i]}\bm{A}_2\bm{M}^{[i]}$, and the initial value $\bm{M}^{[0]} =\bm{A}_4^{-1}\bm{A}_3$. After solving $\bm M$ from~\eqref{eq42}, matrix $\bm N$ can be retrieved by solving the following Sylvester equation
\begin{equation}\label{eq43}
	\bm{N}^{[i]}\bm{E}_1^{[i]} + \bm{E}_2^{[i]}\bm{N}^{[i]} = \bm{A}_2
\end{equation}
Choosing $\epsilon=1/30$, from~\eqref{eq42} and~\eqref{eq43}, we can solve that
\begin{equation*}
	\begin{split}
	\bm{M} = 
	\left[\begin{matrix}
	0.0992 & -0.0334\\
	-2.2051 & -0.0356\\
	\end{matrix}\right]
	, \quad
	\bm{N} =
	\left[\begin{matrix}
	0.0221 & 0.0190 \\
	0.9225 & -0.1621\\
	\end{matrix}\right] .
	\end{split}
\end{equation*}
The pure-fast subsystems~\eqref{eq8} is
\begin{equation*}
	\bm{A}_f = \left[\begin{matrix}
	-0.8401 & 0.9989\\
	-4.7974 & -0.4912\\
	\end{matrix}\right] ,
	\qquad
	\bm{B}_f = \left[\begin{matrix}
	-0.1101 & 6.9565 \times 10^{-5}\\
	-8.6980 & -0.0016\\
	\end{matrix}\right]
\end{equation*}
The pure-slow subsystems~\eqref{eq12} becomes
\begin{equation*}
\bm{A}_s= \left[\begin{matrix}
-0.0149 & -0.0805\\
-4.7974 & -0.4912\\
\end{matrix}\right] ,
\qquad
\bm{B}_s = \left[\begin{matrix}
0.1666 & 0.0008\\
-1.3085 & -0.0003\\
\end{matrix}\right]
\end{equation*}
It can be easily verified that eigenvalues of the subsystems are almost identical with the eigenvalues of the original system. Since the eigenvalues of matrix $\bm{\mathcal{L}}+\bm{\mathcal{B}}$ are $\lambda_1 = \lambda_2 = 1$ and $\lambda_3 = 2$, using criterion~\eqref{eq17}, synchronization can be realized if we choose $c=0.5$ in controller~\eqref{eq21}. Let $\bm Q = \bm I$ and $R=0.001 \cdot \bm I$, and the feedback gain matrices $\bm K_f$ and $\bm K_s$ can be solved from Riccati equation~\eqref{eq20}
\begin{equation*}
	\bm{K}_f = \left[\begin{matrix}
	-15.0567 & -31.4258\\
	0.0410 & -0.0063\\ 
	\end{matrix}
	\right]
	, \quad
	\bm{K}_s
	=\left[\begin{matrix}
	28.8604 & -28.2527\\
	7.2949 & 0.9046\\
	\end{matrix}
	\right] .
\end{equation*}
Suppose that the initial states of the leader is $\bm x^{(0)}(0)=(0, 1, 0, 0.5)^T$ and the initial states of aircrafts 1, 2, and 3 are respectively $\bm x^{(1)}(0)=(0,-0.5,0,1)^T$, $\bm x^{(2)}(0)=(0,2.5,0,0)^T$, and $\bm{x}^{(3)}(0)=(0,0,0,0)^T$. The angles of attack $\alpha$ and pitch angles $\theta$ of leader and followers coordinated by the continuous synchronization protocol are given in~\hyperref[fig3]{Figure 3}. The tracking errors of these attitude parameters $\alpha^{(0)}-\alpha^{(i)}$ and $\theta^{(0)}-\theta^{(i)}$ with $i\in\bm{\mathcal{N}}$ are shown in~\hyperref[fig4]{Figure 4}. From these figures, we can see that the continuous synchronization protocol~\eqref{eq21} can asymptotically synchronize the followers' states with the leader's.

\subsection{Discrete Flight Control System}
\begin{figure}[htpb]
	\centering
	\includegraphics[width=0.58\textwidth]{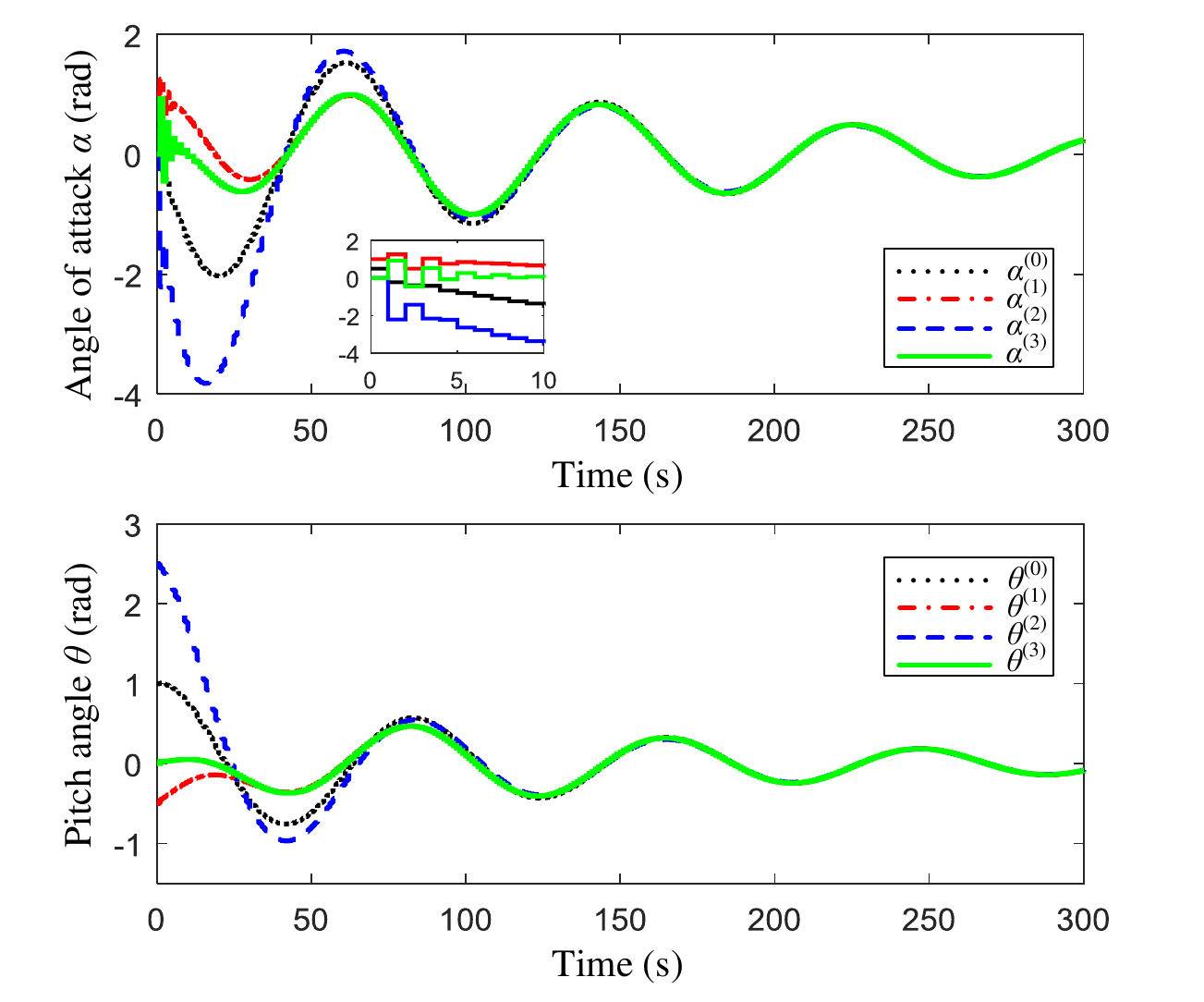}
	\caption{\label{fig5} Longitudinal attitude of aircrafts.}
\end{figure}

\begin{figure}[htpb]
	\centering
	\includegraphics[width=0.58\textwidth]{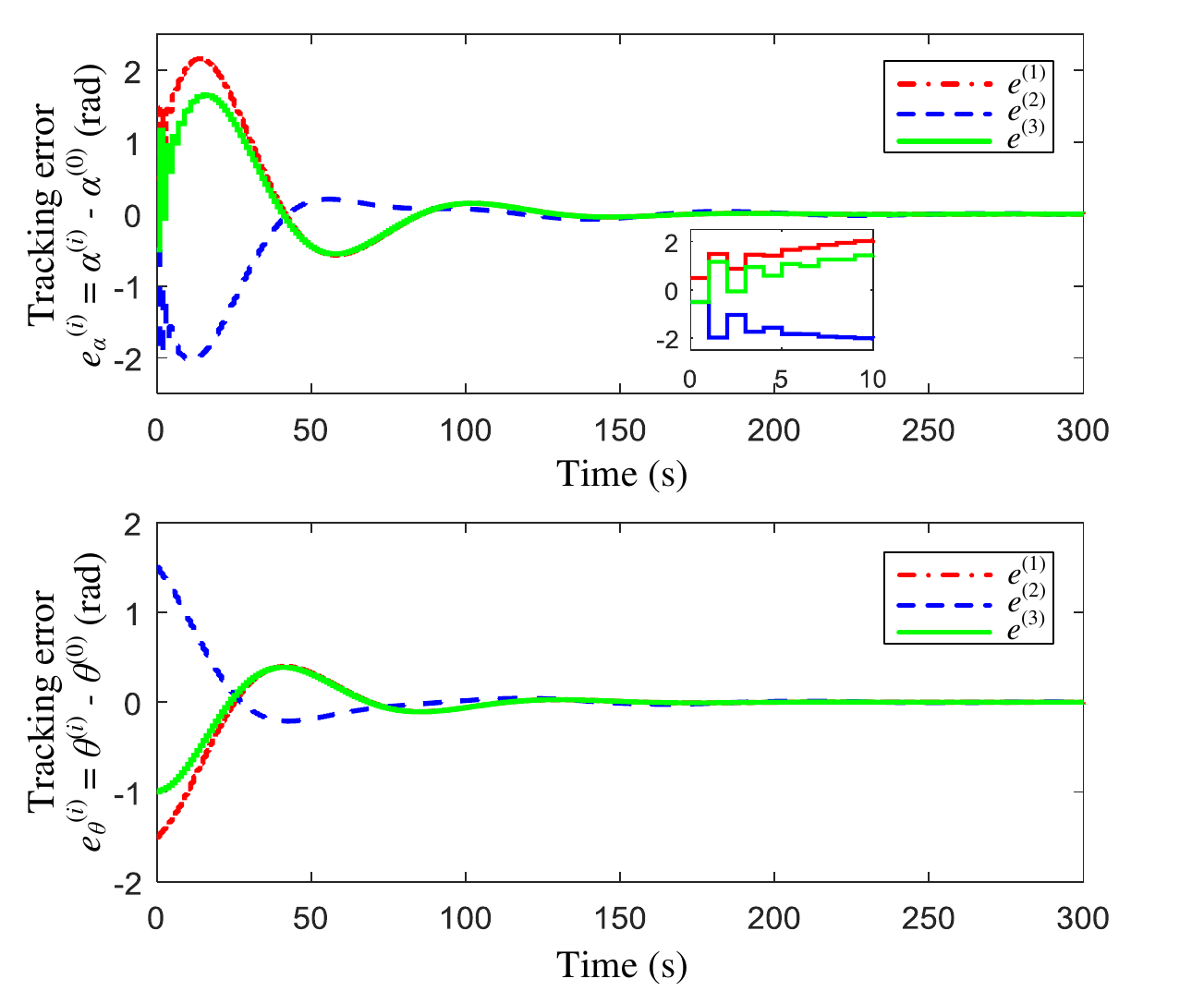}
	\caption{\label{fig6} Tracking error of longitudinal attitude.}
\end{figure}

Sampling the continuous model~\eqref{eq41} with period $T=1s$, we obtain the following discrete model of flight control system, which has the similar form with discrete singularly perturbed model~\eqref{eq3}
\begin{equation}\label{eq44}
	\dot{\bm{x}}(k+1) = \bm{A}_d\bm{x}(k) + \bm{B}_d\bm{u}(k)
\end{equation}
where
\begin{equation*}
	\bm{A}_d = \left[\begin{matrix}
	0.9847 & -0.0799 & 9.054 \times 10^{-4} & -1.076 \times 10^{-3}\\
	0.04159 & 0.9990 & -0.03586 & 0.01268\\
	-0.5466 & 0.04492 & -0.3299 & 0.1932 \\
	2.662 & -0.1004 & -0.9245 & -0.2633\\
	\end{matrix}\right]
	, \quad
	\bm{B} = \left[\begin{matrix}
	0.002893 & 7.361 \times 10^{-4}\\
	-0.08706 & 9.341 \times 10^{-6}\\
	-1.984 & -4.138 \times 10^{-4}\\
	-3.194 & 9.254 \times 10^{-4}\\
	\end{matrix}\right] .
\end{equation*}
$\bm{x}(k)=\left[u,\theta,q,\alpha\right]^T$, $\bm{u}(k)=\left[\delta_e,\delta_T\right]^T$, and $n_1=n_2=2$. Matrices $\bm{M}$ and $\bm{N}$ in~\eqref{eq23} and~\eqref{eq27} can be solved from equations~\eqref{eq26} and~\eqref{eq30} by using the following Newton algorithm:
\begin{equation}\label{eq45}
	\bm{E}_1^{[i]}\bm{M}^{[i+1]} + \bm{M}^{[i+1]}\bm{E}_2^{[i]} = \bm{F}^{[i]}
\end{equation}
where for the discrete case $\bm{E}_1^{[i]} =\bm{A}_4 - \bm{I} + \epsilon\bm{M}^{[i]}\bm{A}_2$, $\bm{E}_2^{[i]} =-\epsilon(\bm{A}_1-\bm{A}_2\bm{M}^{[i]})$, $\bm{F}^{[i]} =\bm{A}_3+\epsilon\bm{M}^{[i]}\bm{A}_2\bm{M}^[i]$, and the initial value $\bm{M}^{[0]} =(\bm{A}_4 - \bm{I})^{-1}\bm{A}_3$. After solving $\bm{M}$ from~\eqref{eq45}, matrix $\bm{N}$ can be retrieved by solving the following Sylvester equation
\begin{equation}\label{eq46}
	\bm{N}^{[i]}\bm{E}^{[i]}_1 + \bm{E}_2^{[i]}\bm{N}^{[i]} = \bm{A}_2
\end{equation}
Choosing $\epsilon=1/30$, from~\eqref{eq45} and~\eqref{eq46}, for discrete case we can solve that
\begin{equation*}
	\bm{M} = \left[\begin{matrix}
	0.0938 & -0.0334\\
	-2.2051 & -0.0356\\
	\end{matrix}\right]
	, \quad
	\bm{N} = 
	\left[\begin{matrix}
	0.0221 & 0.0190\\
	0.9225 & -0.1621\\
	\end{matrix}\right] .
\end{equation*}
Therefore, the discrete pure-fast subsystem~\eqref{eq25} can be described by 
\begin{equation*}
	\bm{A}_f = \left[\begin{matrix}
	-0.3286 & 0.1927\\
	-0.9253 & -0.2613\\
	\end{matrix}\right]
	, \quad
	\bm{B}_f = \left[\begin{matrix}
	-1.8885 & 0.0016\\
	-3.2983 & -0.0478\\
	\end{matrix}\right].
\end{equation*}
The discrete pure-slow subsystem~\eqref{eq29} can be described by 
\begin{equation*}
	\bm{A}_s = \left[\begin{matrix}
	0.9823 & -0.0799\\
	0.0729 & 0.9982\\
	\end{matrix}\right]
	, \quad
	\bm{B}_f = \left[\begin{matrix}
	-1.8885 & 0.0016\\
	-3.2983 & -0.0478\\
	\end{matrix}
	\right].
\end{equation*}
We can easily verify that the eigenvalues of the subsystems are almost identical with the original system. Let $\bm{Q}=\bm{I}$, and the feedback gain matrices $\bm{K}_f$ and $\bm{K}_s$ can be solved from the discrete Riccati equation~\eqref{eq35}
\begin{equation*}
	\bm{K}_f = \left[\begin{matrix}
	0.1801 & -0.0917\\
	6.9349 & 11.8009\\
	\end{matrix}\right]
	, \quad
	\bm{K}_s = \left[\begin{matrix}
	-0.0289 & -0.3882\\
	44.5447 & -1.9819\\
	\end{matrix}\right].
\end{equation*}
The radius defined in~\eqref{eq37} of pure-fast and pure-slow systems are respectively $r_f=1.001$ and $r_s=0.9981$. The eigenvalues of the weighted graph matrix $\bm{\Gamma}$ associated with the graph in~\hyperref[fig2]{Figure 2} are $0.5$, $0.5$, and $2/3$, which can be contained in a covering circle centered at  $(1/c,0)=(7/12,0)$ with radius $r_0=1/12$. Therefore, it is reasonable for us to choose $c_s=c_f=12/7$ in~\eqref{eq40}. Assume that the initial states of leader and followers are the same, and the sampling period $T=1s$. Then angles of attack $\alpha$ and pitch angles $\theta$ coordinated by the discrete synchronization protocol~\eqref{eq39} are given in~\hyperref[fig5]{Figure 5}. The attitude tracking errors $\alpha^{(0)}-\alpha^{(i)}$ and $\theta^{(0)}-\theta^{(i)}$ with $i\in\bm{\mathcal{N}}$ are shown in~\hyperref[fig6]{Figure 6}.

With the same initial states, command signals from the leader agent are almost identical in~\hyperref[fig3]{Figure~3} and~\hyperref[fig5]{Figure~5}, and the states of different aircrafts synchronize with the states of leader via the discrete decentralized tracking protocol\eqref{eq39}. \hyperref[fig6]{Figure~6} tells that tracking error $\bm x^{(i)}(k)-\bm x^{(0)}(k)$ asymptotically converges to zero, which corroborate the effectiveness of the decentralized control scheme.

\section{Conclusion}\label{sec6}
Decentralized synchronization control problems for both continuous and discrete singularly perturbed systems with fast and slow scales have been investigated in this research paper. With SPaTS technique, the original systems were decomposed into pure-slow and pure-fast subsystems. Locally optimal synchronization controllers were synthesized respectively to asymptotically stabilize each subsystems. The overall controller for each agent is the superposition of sub-controllers. The effectiveness of the control schemes was verified with flight control systems for aircraft formation flying.

\section*{Acknowledge}
The authors specially acknowledge the staff and readers from arXiv.org.


\begin{thebibliography}{}
\bibitem{Negenborn_CSM_2014}
R.R. Negenborn, and J.M. Maestre, "Distributed model predictive control: An overview and roadmap of future research opportunities", IEEE Control System Magazine, vol. 34, no. 4, pp. 87-97 (2014).
\bibitem{Wang_TAC_2013}
X. Wang, and N. Hovakimyan, "Distributed control of uncertain networked systems: A decoupled design", IEEE Transactions on Automatic Control, vol. 58, no. 10, pp. 2536-2549 (2013).
\bibitem{Cao_TII_2013}
Y. Cao, W. Yu, W. Ren, and G. Chen, "An  overview of recent progress in the study of distributed multi-agent coordination", IEEE Transactions on Industrial Informatics, vol. 9, no. 1, pp. 427-438 (2013). 
\bibitem{Ren_2008}
W. Ren, and R.W. Beard, \textit{Distributed consensus in multi-vehicle cooperative control} (Springer, London, 2008).
\bibitem{Saber_PIEEE_2007}
R. Olfati-Saber, J.A. Fax, and R.M. Murray, "Consensus and cooperation in networked multi-agent systems", Proceedings of the IEEE, vol. 95, no. 1, pp. 215-233 (2007).
\bibitem{Lu_TNNLS_2012}
J. Lu, J. Kurths, J. Cao, N. Mahdavi, and C. Huang, "Synchronization control for nonlinear stochastic dynamical networks: Pinning impulsive strategy", IEEE Transactions on Neural Networks and Learning Systems, vol. 23, no. 2, pp. 285-292 (2012).
\bibitem{Lewis_2013}
F.L. Lewis, H. Zhang, K. Hengster-Movric, and A. Das, \textit{Cooperative control of multi-agent systems} (Springer, New York, 2013).
\bibitem{Su_TAC_2016}
S. Su, and Z. Lin, "Distributed consensus control of multi-agent systems with higher order agent dynamics and dynamically changing interaction topologies", IEEE Transactions on Automatic Control, vol. 61, no. 2, pp. 515-519 (2016).
\bibitem{Pilloni_TAC_2016}
A. Pilloni, A. Pisano, Y. Orlov, and E. Usai, "Consensus-based control for a network of diffusion PDEs with boundary local interaction", IEEE Transactions on Automatic Control, vol. 61, no. 9, pp. 2708-2713 (2016).
\bibitem{Tong_TSMC_2016}
S. Tong, L. Zhang, and Y. Li, "Observed-based adaptive fuzzy decentralized tracking control for switched uncertain nonlinear large-scale systems with dead zones", IEEE Transactions on Systems, Man, and Cybernetics: System, vol. 46, no. 1, pp. 37-47 (2016).
\bibitem{Naidu_JGCD_2001}
D.S. Naidu, and A.J. Calise, "Singular perturbations and time scales in guidance and control of aerospace systems: A survey", Journal of Guidance, Control, and Dynamics, vol. 24, no. 6, pp. 1057-1078 (2001).
\bibitem{Gajic_2001}
Z. Gajic, and M.T. Lim, \textit{Optimal control of singularly perturbed linear systems and applications} (Marcel Dekker, New York, 2001).
\bibitem{Naidu_2002}
D.S. Naidu, "Singular perturbations and time scales in control theory and applications: An overview", Dynamics of Continuous, Discrete and Impulsive Systems Series B: Applications \& Algorithm, vol. 9, pp. 233-278 (2002).
\bibitem{Zhang_IJISS_2014}
Y. Zhang, D.S. Naidu, C. Cai, and Y. Zou, "Singular perturbations and time scales in control theory and applications: An overview 2002-2012", International Journal of Information Systems Sciences, vol. 9, pp. 1-36 (2014).
\bibitem{Zhang_WSEAS-TSC_2014}
Y. Zhang, H. Nguyen, D.S. Naidu, Y. Zou, and C. Cai, "Time scale analysis and synthesis for model predictive control", WSEAS Transactions on Systems and Control, vol. 9, pp. 130-139 (2014).
\bibitem{Litkouhi_TAC_1985}
B. Litkouhi, and H. Khalil, "Multirate time and composite control of two-time-scale discrete-time systems", IEEE Transactions on Automatic Control, vol. 30, no. 7, pp. 645-651 (1985). 
\bibitem{Zhang_IJSS_2016}
Y. Zhang, D.S. Naidu, C. Cai, and Y. Zou, "Composite control of a class of nonlinear singularly perturbed discrete-time systems via D-SDRE", International Journal of Systems Science, vol. 47, no. 11, pp. 2632-2641 (2016).
\bibitem{Chang_SIAM-JMA_1972}
K.W. Chang, "Singular perturbations of a general boundary value problem", SIAM Journal on Mathematical Analysis, vol. 3, no. 3, pp. 520-526 (1972).
\bibitem{Zhang_TAC_2011}
H. Zhang, F.L. Lewis, and A. Das, "Optimal design for synchronization of cooperative systems: state feedback, observer and output feedback", IEEE Transactions on Automatic Control, vol. 56, no. 8, pp. 1948-1952 (2011).
\bibitem{Naidu_NASA_1988}
D.S. Naidu, and D.B. Price, "Singular perturbations and time scales in the design of digital flight control systems", NASA Technical Paper 2844 (1988).
\bibitem{Naidu_1988}
D.S. Naidu, \textit{Singular perturbation methodology in control systems} (IET, London, 1988).
\bibitem{Elliott_TAC_1977}
J.R. Elliott, "NASA’s advanced control law program for the F-8 digital fly-by-wire aircraft", IEEE Transactions on Automatic Control, vol. 22, no. 5, pp. 753-757 (1977).





% Format for books
%\bibitem{RefB}
%Book Author, \textit{Book title} (Publisher, place, year) page numbers
% etc
\end{thebibliography}
\end{document}